\documentclass[11pt,dvips]{article}
\usepackage{epsfig,times} 
\usepackage{picinpar}
\usepackage{wrapfig}
\usepackage{floatflt}
\makeatletter
\renewcommand{\section}{\@startsection{section}{1}{0in}
	{0.4\baselineskip}{0.1\baselineskip}{\Large\bf}}
\renewcommand{\subsection}{\@startsection{subsection}{2}{0in}
	{0.25\baselineskip}{-\baselineskip}{\large\bf}}
\renewcommand{\subsubsection}{\@startsection{subsubsection}{3}{0in}
	{0.1\baselineskip}{-\baselineskip}{\normalsize\bf}}
\makeatother
\begin{document}

%
\thispagestyle{myheadings}
%
\markright{OG.4.3.04}
\begin{center}
%
{\LARGE \bf Construction of New 7m Imaging Air \v Cerenkov Telescope 
of CANGAROO }
\end{center}

\begin{center}
%
%
{\bf
T.~Tanimori\footnote[1]{
Graduate School of Science and Engineering, Tokyo Institute of Technology, 
Meguro, Tokyo, 152-8551, Japan,
$^{2}$Department of Physics and Mathematical Physics, University of 
   Adelaide, South Australia 5005, Australia, 
$^{3}$Institute of Space and Astronautical Science,
   Sagamihara, Kanagawa 229-8510, Japan, 
$^{4}$Department of Physics, Yamagata University, 
Yamagata 990-8560, Japan, 
$^{5}$Faculty of Management Information, Yamanashi Gakuin Univeristy,  Kofu, 
Yamanashi 400-8575, Japan, 
$^{6}$Department of Physics, Tokai University, 
 Hiratsuka, Kanagawa 259-1292, Japan, 
$^{7}$Institute for Cosmic Ray Research, University of Tokyo,
     Tanashi, Tokyo 188-8502, Japan, 
$^{8}$STE Laboratory, Nagoya University,
   Nagoya, Aichi 464-860, Japan, 
$^{9}$National Astronomical Observatory, Tokyo 181-8588, Japan, 
$^{10}$Faculty of Science, Ibaraki Univeristy, 
   Mito, Ibaraki 310-8521, Japan, 
$^{11}$LPNHE, Ecole Polytechnique. Palaiseau CEDEX 91128, France,
$^{12}$Institute of Physical and Chemical Research,
   Computational Science Laboratory, Institute of Physical and Chemical
   Research, Wako, Saitama 351-0198, Japan,
$^{13}$Faculty of Engineering, Kanagawa University,
 Yokohama, Kanagawa 221-8686, Japan

}, 
S.A.~Dazeley$^2$,
P.G.~Edwards$^3$,
 S.~Gunji$^4$, S.~Hara$^1$,  
T.~Hara$^5$, J.~Jinbo$^6$, 
A.~Kawachi$^7$, T.~Kifune$^7$, 
H.~Kubo$^1$, 
J.~Kushida$^1$, Y.~Matsubara$^8$, 
Y.~Mizumoto$^9$,  
M.~Moriya$^1$, M.~Mori$^7$, 
H.~Muraishi$^{10}$, Y.~Muraki$^8$, 
T.~Naito$^5$, K.~Nishijima$^6$,  
J.R.~Patterson$^2$, M.D.~Roberts$^7$, 
G.P.~Rowell$^7$, T.~Sako$^{8,11}$, 
K.~Sakurazawa$^1$, Y.Sato$^7$,
R.~Susukita$^{12}$, 
T.~Tamura$^{13}$, S.~Yanagita$^{10}$, 
T.~Yoshida$^{10}$, T.~Yoshikoshi$^7$, 
A.~Yuki$^8$  
}
\end{center}

\begin{center}
{\large \bf Abstract\\}
\end{center}
\vspace{-0.5ex}
CANGAROO group has constructed the new  
large imaging Air \v Cerenkov telescope to exploit hundred GeV region
gamma-ray astronomy in March 1999 at Woomera, South Australia.
It has a 7m parabolic mirror consisting of 60 small plastic spherical mirrors,
and  a fine imaging camera with 512 PMTs covering the field of view of 3 degree.  
Observation will  start from  July 1999. 
\vspace{1ex}

%
%
\section{Introduction:}
\label{intro.sec}

In this decade  Very High Energy(VHE) gamma-ray astronomy 
has been dramatically 
advanced due to the appearance of an imaging Air \v Cerenkov Telescope,
and about ten  TeV gamma-ray emitters have been found so far.
Nowadays VHE gamma-ray astronomy is just proceeding to the next stage
in exploiting  hundred GeV or sub-hundred GeV region.
For example Whipple group promotes the construction of
the array of  eight 10m imaging telescopes.
German groups also propose two main projects: HESS consisting of
16 X 10m imaging telescope and MAGIC of one big 17m imaging telescope looking 
for several ten GeV gamma rays. 
In this manner, new stage of VHE astronomy  will  
certainly begin at the opening of 21st century. 

CANGAROO group has also pushed  ahead the next project consisting of
four 10m imaging telescopes.
In 1995 the construction of one imaging telescope was approved (CANGAROO II project).
Due to the limit of the approved fund,
we decided to construct the telescope with a 7m mirror,
of which frame and base however can 
sustain a large mirror of a 10m diameter.
Therefore, in near future, 
this telescope will be easily extended to the 10m mirror
by adding more  mirrors.

\begin{figure}[h]
\begin{center}
\label{fig:tele}
\epsfig{file=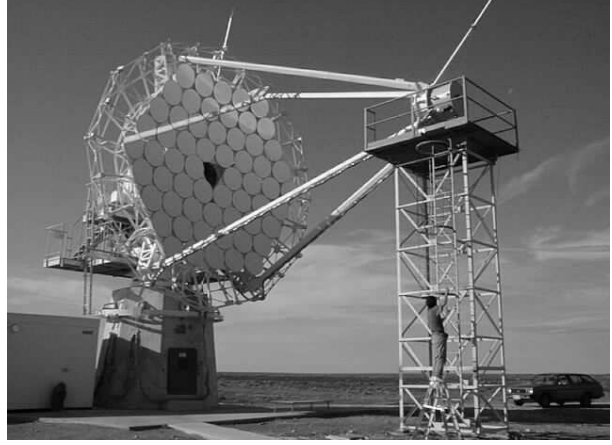, width=8cm}
\caption{ Photo image of new CANGAROO 7m telescope at Woomera
}
\end{center}
\end{figure}

In December 1998, all components of the telescope and the gamma camera
were prepared in Japan and shipped to Australia.
In the middle of March 1999, the telescope construction was
completed  at Woomera
as shown in Fig.\ref{fig:tele}.
After then the short observation 
for studying the performances of the telescope was done.
Here we present the overview of this new telescope
and the brief report of its performance as an 
imaging \v Cerenkov telescope.

\section{Performance of Mirror:}
\label{mirror.sec}

The design concept and features of the 7m telescope were already presented in 
Tanimori et al.\ 1995 and Matsubara et al.\ 1997.
We adopted a  7m parabolic mirror with a 8m focal length, which
consists of sixty spherical-composite mirrors with a 80cm diameter
(total light-collection area of 30.1 m$^2$),
in order to keep the duration of the arrival times of  \v Cerenkov photons 
within a few ns.
For the extension to a 10m mirror,
an additional fifty-four mirrors will be  attached on the outer frame around 
the present mirrors.
One of  challenges in this telescope was to use plastic 
as a material of the composite mirror.
The mirror consists of mainly Carbon Fiber Reinforced Plastic(CFRP) sheets and thin 
aluminum foils.
By our long endeavor, the accuracy of the mirror surface was achieved
within about 20 $\mu$m, which corresponds to the blur of 0$^{\circ}$.08(FWHM)
for a parallel beam.
The weight of this mirror is only one fifth of that made by glass.
The reflectivity of this mirror can be kept $\sim 80$\%  in several years 
by washing its surface by water.
The blur of the 7m parabolic  mirror consisting of 60 small plastic mirrors
was measured using star images taken by the CCD camera, 
and obtained to be $\sim$  0$^{\circ}.15$(FWHM).
This resolution  is within an allowable range, but not satisfactory, 
considering the fine resolution of the imaging camera.
The blurs of about 10 mirrors were found to be much worse than others. 
It will be improved by replacing those bad mirrors
in next year.
The details about the performances of the 7m mirror and
the composite mirror are presented in Kawachi et al.\ 1999.

Also using star images,
the tracking accuracy  and the deformation of the 7m mirror due to 
the camera weight were estimated by changing
the azimuthal and zenith angles.
The results obtained were 0$^{\circ}$.006 and  0$^{\circ}$.007 respectively,
which are much  smaller than a blur of the mirror.

\section{Performance of Camera \& Electronics:}
\label{camera.sec}

Figure 2 shows the front view of the camera attached in the focal plane,
where that of 3.8m telescope is also presented.
This camera consists of 512 pixels to covers a field of view (FOV)
of diameter $\sim$ 3$^{\circ}$,
Each pixel covers 0$^{\circ}$.115 $\times$ 0$^{\circ}$.115 (16mm $\times$ 16mm),
and 13 mm $\phi$ photomultiplier (PMT: Hamamatsu R4124UV) was used as a pixel detector.
The photocathode of this PMT has an area of 10 mm$\phi$ and cover about 35\% of the FOV. 
The Array of hollow light collectors were attached in front of the PMTs 
in order to increase the collection area of the camera by twice. 
Sixteen PMTs are housed in one module unit with a common bleeder circuit.
PMTs are operated with a low gain of $\sim$ 10$^5$ 
to avoid the long-gain drop (more than a few ten minutes) due to the passage of bright
stars.  

\begin{figure}[h]
\begin{center}
\label{fig:camera}
\epsfig{file=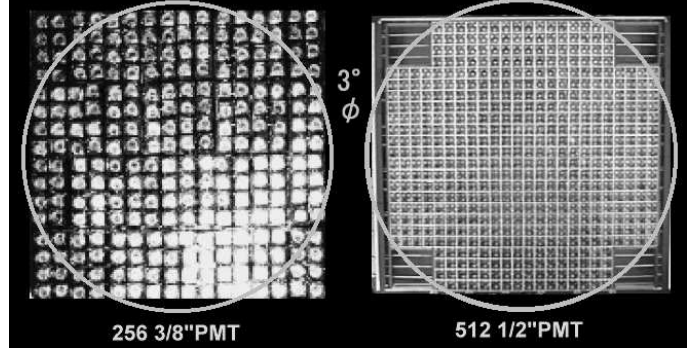, width=9cm}
\caption{ Front view of both the cameras of the 3.8 telescope (left) and 
the new 7m telescope (right). 
}
\end{center}
\end{figure}

Buffer amplifiers (LeCroy TRA402) are also installed in the module box to sustain 
the total gain of $\sim$ 10$^7$ (after amplification) and to feed signals through long
twisted cables(36 m). 
The whole camera consists of 32 module units.

Signals were fed to electronics circuits in the hut located by the base of the telescope.
Here all timings and pulse heights of hit PMTs are measured, where a hit PMT
means that its  pulse become larger than  the preset value
(usually three or four photoelectrons).
The detail is described in Mori et al. 1999.
Triggers are generated when both the number of a hit PMT (H-sum) 
and the linear sum of all PMT signals (L-sum)
exceed the preset values.
Inner-most 16  of 32 modules  are concerned with the trigger,
which covers a $\sim 1^{\circ}$.8 diameter of the FOV. 
The observation for studying  the camera and electronics 
was done during few days in March. 
When requiring the H-sum $\ge \sim 4$  and L-sum $\ge \sim 1.2 \times$ of the linear
sum of the night sky
background fluctuation,
triggers were generated at $\sim$ 11 Hz.

In this condition,
the timing distribution of hit PMTs accumulated 
for all events are shown in Fig.3,
where no correction for the time jitter was not applied. 
Timings of hit PMTs almost concentrates within 50ns ( 1bin = 0.5 ns),
which indicates that  more than 90\% of events are triggered not by an accidental
coincidence due to the night sky background but by a muon or a shower.

\begin{figure}[h]
\begin{center}
\label{fig:timing}
\epsfig{file=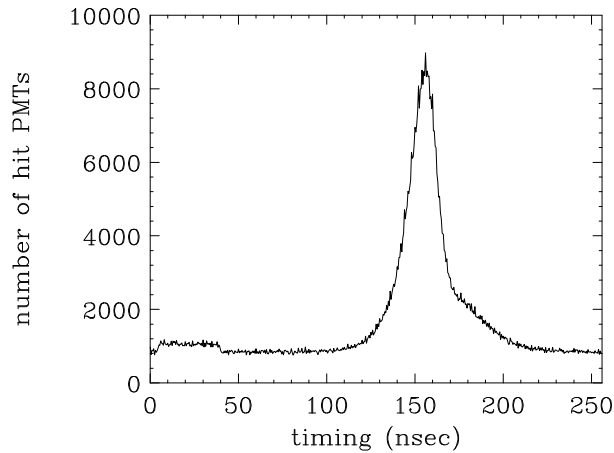, width=8cm}
\caption{ Plot of arrival timings of all of hit PMTs
}
\end{center}
\end{figure}

Figure 4 shows the distribution of the L-sum for those events.
Note that most events including in this figure  are not accidental but physical
events.     
The higher part of the distribution ($\sim$ 45\% )
clearly corresponds to those triggered by showers (power law).
Note that the slope becomes flatter as energy decreasing.
This might be due to the decrease of the detection efficiency  
at the trigger level for hadrons below $\sim$1 TeV,
which was expected for the large telescope having a fine imaging pixels
from the simulation study (Aharonian et al. 1994; Tanimori et al. 1994). 
On the other hand, the dominant peak near 400 counts is considered  
due to single muons
from the narrow-timing concentration  and the hit pattern on the camera.
In 3.8m telescope, muon events were rarely detected since its detectable
energy was relatively high ($\sim$ 1.5 TeV).
We are now estimating the threshold energy by comparing the simulation
for hadrons and muons.

\begin{figure}[h]
\begin{center}
\label{fig:spec}
\epsfig{file=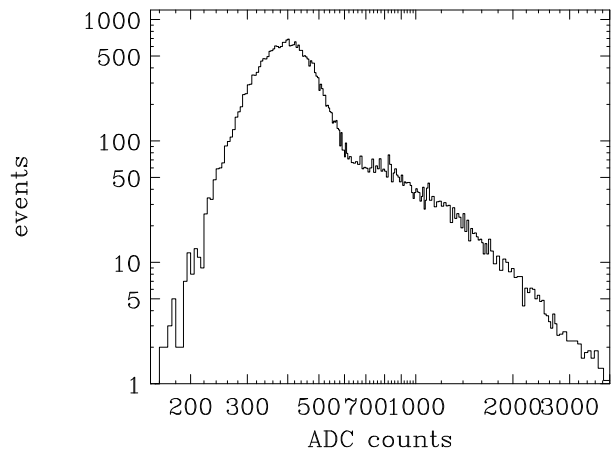, width=8cm}
\caption{ Distribution of the linear sum of all PMTs for triggered events
}
\end{center}
\end{figure}

\section{Summary:}
\label{summ.sec}

The construction of New CANGAROO 7m telescope has been completed in March,
and the telescope has detected many  shower and muon events.
Now the detailed study of the trigger condition is ongoing.
In May and June, the tuning of electronics and trigger condition
is being carried out.
In particular the each L-sum of a module nuit including 
will be individually discriminated from the L-sum due to the night sky background
in one module unit,
by which the threshold energy will be expected to be decreased.
Then we will start the normal observation from July.
Also the fast pattern trigger using the hit pattern of 32 module units will be
applied late of this year.

Finally  we are now about to start the  construction of another three
10m imaging telescopes (CANGAROO III project), 
which has been approved by Japanese government in April 1999.
CANGAROO III includes the extension of the 7m telescope to 10m. 
The array of four  10m imaging telescopes for the stereo observation will 
be completed by  2003 at Woomera.

\vspace{1ex}
\begin{center}
{\Large\bf References}
\end{center}
%
Aharonian, F.\ et al., 1994, Proc.\ Towards a Major Atmospheric Cherenkov Detector III(Tokyo, 1994)\\ 
Kawachi, A.\ et al., 1999, Proc. 26th ICRC (Salt Lake City, 1999)\\
Matsubara, Y.\ et al., Proc.\ 26th ICRC (Durban, 1997) \\ 
Mori, M.\ et al., 1999, Proc. 26th ICRC (Salt Lake City, 1999)\\
Tanimori, T.\ et al., 1994, Proc.\ Towards a Major Atmospheric Cherenkov Detector III(Tokyo, 1994)\\ 
Tanimori, T.\ et al., 1995, Proc.\ Towards a Major Atmospheric Cherenkov Detector IV(Padova, 1995)\\ 
\end{document}